\documentclass[aps,prl,preprint,groupedaddress]{revtex4-1}
\usepackage{braket}
\usepackage{amsmath}
\usepackage{graphicx}
\usepackage{color}

\begin{document}

\author{Mit H. Naik and Manish Jain}
%\altaffiliation{A shared footnote}

\affiliation{Center for Condensed Matter Theory, Department of Physics, Indian Institute of Science, Bangalore 560012, India}
%\altaffiliation{Current address: Some other place, Othert\"own,
%Germany}
%\author{I. Ken Groupleader}
%\altaffiliation{A shared footnote}
%\email{mjain@iisc.ac.in}
%\phone{+123 (0)123 4445556}
%\fax{+123 (0)123 4445557}
%\alsoaffiliation[Second University]
%{Department of Chemistry, Second University, Nearby Town}
%\author{Susanne K. Laborator}
%\email{s.k.laborator@bigpharma.co}
%\affiliation[BigPharma]
%{Lead Discovery, BigPharma, Big Town, USA}
%\author{Kay T. Finally}
%\affiliation[Unknown University]
%{Department of Chemistry, Unknown University, Unknown Town}
%\alsoaffiliation[Second University]
%{Department of Chemistry, Second University, Nearby Town}

%%%%%%%%%%%%%%%%%%%%%%%%%%%%%%%%%%%%%%%%%%%%%%%%%%%%%%%%%%%%%%%%%%%%%
%% The document title should be given as usual. Some journals require
%% a running title from the author: this should be supplied as an
%% optional argument to \title.
%%%%%%%%%%%%%%%%%%%%%%%%%%%%%%%%%%%%%%%%%%%%%%%%%%%%%%%%%%%%%%%%%%%%%
\title
  {Substrate screening effects on the quasiparticle band gap and
defect charge transition levels in MoS$_2$ }

%%%%%%%%%%%%%%%%%%%%%%%%%%%%%%%%%%%%%%%%%%%%%%%%%%%%%%%%%%%%%%%%%%%%%
%% Some journals require a list of abbreviations or keywords to be
%% supplied. These should be set up here, and will be printed after
%% the title and author information, if needed.
%%%%%%%%%%%%%%%%%%%%%%%%%%%%%%%%%%%%%%%%%%%%%%%%%%%%%%%%%%%%%%%%%%%%%
%\abbreviations{IR,NMR,UV}

\begin{abstract}
% insert abstract here
Monolayer MoS$_2$ has emerged as an interesting material for
nanoelectronic and optoelectronic devices.
The effect of substrate screening and defects on the electronic structure
of MoS$_2$ are important considerations in the design of such devices.
We find a giant renormalization to the free-standing quasiparticle band gap
in the presence of metallic substrates,
in agreement with recent scanning tunneling spectroscopy and photoluminescence
experiments.
Our sulfur vacancy defect calculations using the DFT+GW formalism,
reveal two CTLs in the pristine band gap of MoS$_2$. 
The (0/-1) CTL is significantly renormalized with 
the choice of substrate, with respect to the pristine valence band maximum.
The (+1/0) level, on the other hand, is pinned 100 meV above the pristine VBM 
for the different substrates. This opens up a pathway to effectively engineer
defect charge transition levels in 2D materials through choice of substrate.
\end{abstract}

\maketitle

\section{Introduction}

MoS$_2$, part of the family of the layered transition metal dichalcogenides (TMDC), has
garnered great interest owing to its diverse applications in nanoelectronics and optoelectronics
\cite{PRL_Mak,NL.Splendani,NN.Wang}.
High current on-off ratios in field effect transistors as well as efficient valley
and spin control with optical helicity have been
achieved using MoS$_2$\cite{NN_Radisavljevic,NN_Mak}. The direct band
gap in monolayer MoS$_2$ is exploited in building
ultrasensitive phototransistors \cite{NN_Roy,NN_Lopez,AP_Ye}.
MoS$_2$ is also considered a promising alternative to platinum as a catalyst in the
hydrogen evolution reaction \cite{NM_Li,NL_Voiry,AN_Chang}.

Effect of the dielectric environment and defects on the electronic
structure of MoS$_2$ are the most important considerations
in the design of devices using MoS$_2$
\cite{NC_Hong,AN_Liu,AMI_Addou,NL_Lin}.
Single layer MoS$_2$,  
achieved through transfer
post exfoliation or through direct epitaxial growth,
 \cite{AN_Liu,NL_Zhang}
is typically supported on a substrate.
\cite{NL_Zhang,SR_Man,NL_Hsieh,PRB_Bruix,APL.Ghatak}
Scanning tunneling spectroscopy (STS) measures the 
quasiparticle band gap of MoS$_2$ on metallic
substrates. \cite{NL_Zhang,NComm_Huang} The screening
from the metal is consistently found to reduce the gap.
\cite{NL_Zhang,AN_Zhang,AN_Ulstrup,AN_Liu,NComm_Huang}
In the presence of graphene or graphite as substrate, a renormalization
larger than 300 meV is observed in the quasiparticle band gap of MoS$_2$.
\cite{NL_Zhang,AN_Ulstrup,AN_Liu,JPCC_Thyg,2D_Thyg,NL_Thyg,SR.Ryou}
Photoluminescence properties
of MoS$_2$ are also strongly influenced by the ambient dielectric environment.
\cite{NL_Zhang,PRM_Borghardt}
A red-shift in exciton peaks due to substrate screening effects is computed in 
Ref. \cite{NC.Rohlf}.

The most abundant native defect found in monolayer MoS$_2$ is the sulfur vacancy.
\cite{NC_Hong}
Sulfur vacancies
induce states in the gap of pristine MoS$_2$, thus
affecting its electronic and optical properties.
\cite{NM_Li,PRB_Noh,NL_Zhou}
Noise nanospectroscopy to probe the ionization 
dynamics of sulfur vacancy defects in MoS$_2$ show 0 and -1 
charge states of the defect to be stable. \cite{NC.Song}
Charged-impurity 
scattering from sulfur vacancies could thus be an important 
factor limiting the mobility of carriers in MoS$_2$. \cite{NC.Song}
While effort is constantly made at attaining lower concentration
of defects in MoS$_2$, sulfur vacancies have found a
favourable role in enhancing the rate of hydrogen evolution reaction.
\cite{NM_Li} Defect levels induced by sulfur vacancies in the band gap 
are responsible for the adsorption of hydrogen. \cite{NM_Li, CM.Ouyang}
The strength of 
hydrogen binding at the defect sites is determined by the 
difference in energy between the defect state and the Fermi 
level. \cite{CM.Ouyang}
This binding is favourable for catalysis if the hydrogen is 
bound neither too strongly nor too weakly. \cite{NL.Tsai}
Pathways to engineer the position of the defect level in the gap 
are thus vital to enhance the hydrogen evolution reaction. \cite{NM_Li,NL.Ye}

A number of theoretical calculations, based on first principles 
density functional theory (DFT),
\cite{PRB.Kohn,PR.Hohenberg}
have been carried out to study sulfur vacancies in monolayer MoS$_2$.
\cite{PRB_Noh,PRB_Komsa2,PRB.Sensoy,JPCC_Talat,IN_Santosh,PRB_Sanyal}
Calculation of charged defects
and in turn the charge transition levels (CTLs) within DFT
has pitfalls owing to the underestimation of the band gap in Kohn-Sham DFT.
\cite{PRL_Perdew}
Some simulations use hybrid functionals, as
proposed by Heyd, Scusseria and Erzernhof (HSE), \cite{JCP_HSE}
as an attempt to overcome the band gap problem \cite{PRB_Noh,PRB_Komsa2}.
However, the band gap of monolayer MoS$_2$ computed using HSE is
about 2.2 eV,
\cite{PRB_Noh,PRB_Komsa2,APL_Jason,PRB_Bhatta}
which is 0.5 eV smaller
than the experimentally measured quasiparticle band gap of
free-standing MoS$_2$. \cite{NL.Krane}
%computed using many body
%perturbation theory (MBPT) within the GW approximation.
%\cite{hedin70,hedin65,PRL.Qiu,PRL.Erratum.Qiu}
This could affect the results on the
defect CTLs in MoS$_2$. Furthermore, substrate screening effects 
can not be effectively studied using DFT or hybrid functionals.
\cite{PRL_Jain_HSE}

Many body perturbation theory in the GW approximation has been 
combined with DFT in the well known DFT+GW
formalism \cite{PRL_Rinke,RMP_Walle} and has been used to predict accurate
defect formation energies and CTLs.
\cite{PRB_Andrei,PRL_Jain,JPCM_Chen}
In the DFT+GW formalism, the energy associated with atomic relaxations 
on adding an electron is taken into account at the DFT level, and the 
quasiparticle excitation energy at the GW level. 
%The 
%DFT+GW formalism incorporates atomic relaxation energy upon adding an electron
%to the system at the DFT level, and
%the quasiparticle excitation energy at the GW level.
Performing GW calculations on transition metal dichalcogenides (TMDCs) in particular 
are computationally challenging owing to the stringent convergence parameters.
\cite{PRL.Qiu,NL.Aaron,PRB.Qiu}
A DFT+GW study on defects in TMDCs, which entail super cell calculations,
require massive computation.
Additionally, the effect of substrate screening can be taken into 
account accurately within the GW approximation. \cite{NMat.Ugeda,NL_Qiu,NL.Aaron,PRL_Jain_HSE}
While it is known that metallic substrates renormalize the pristine 
quasiparticle band gap significantly, it isn't apparent if the defect levels will 
continue to prevail in the pristine gap or be pushed above or below the 
pristine CBM or VBM, respectively. 

In this work, we study the effect of substrate screening on the quasiparticle
band gap and defect charge transition levels in monolayer MoS$_2$.
We have considered graphene, hexagonal BN, graphite and SiO$_2$ as substrates.
At the DFT level, we find that these substrates do not influence the
electronic structure of MoS$_2$. 
This is due to the absence of long range
correlation effects in DFT.
At the GW level, however, we find a significant renormalization in the 
quasiparticle band gap in the presence of these substrates.
The quasiparticle gap is
renormalized from its free-standing value of 2.7 eV to 2.4 eV in the presence of
graphene and to 2.2 eV in the presence of graphite as substrate. In the presence of
BN or SiO$_2$, the gap is close to that of free-standing MoS$_2$. These results are
in good agreement with recent experimental measurements. \cite{ADM_Shi,AN_Liu,NL_Zhang,
APL_Chun,NComm_Huang,NL.Krane,PRL.Yao,NC.Chiu}
We also study the electronic structure of MoS$_2$ in the presence of
sulphur vacancy defects. The sulfur vacancy induces states in the pristine band gap
of MoS$_2$. We compute the CTLs of the sulfur vacancy using
the DFT+GW formalism. Two CTLs appear in the quasiparticle gap:
the (+1/0) and (0/-1) levels are 0.1 eV and 2.2 eV above the pristine valence band maximum (VBM),
respectively.
We further study the effect of substrate screening on the CTLs. The (+1/0) level
lies within 100 meV of the VBM in the presence of substrates as well.
The (0/-1) level, on the other hand, is significantly renormalized and 
can be tuned with the choice of substrate.

\section{Computation details}

The density functional theory (DFT) calculations are performed using the plane-wave
pseudopotential package Quantum Espresso. \cite{QE.Giannozi}
We use the local density approximation for the exchange correlation
functional and norm-conserving pseudopotentials.
The wavefunctions are expanded in plane waves upto an energy
cut off of 250 Ry.
For the unit cell MoS$_2$ calculations, the cell dimension in the out-of-plane direction
is fixed at 35 $\textrm{\AA}$ and the Brillouin zone sampled with a
12$\times$12$\times$1 k-point grid. The relaxed in-plane lattice parameter of 
MoS$_2$ is 3.15 $\textrm{\AA}$.
We simulate a sulfur vacancy in MoS$_2$ by constructing a 5$\times$5
in-plane super cell  and removing a sulfur atom.
The cell dimension in the out-of-plane direction here is fixed at 18 $\textrm{\AA}$.
K-point sampling of 2$\times$2$\times$1 is used in the
super cell calculations.
The formation energy of charged sulfur vacancies computed at the DFT level need to be
corrected for the spurious electrostatic interaction between the charge and
its periodic images. The electrostatic
corrections are computed using the CoFFEE code. \cite{CoFFEE.Naik}

The quasiparticle excitation energies are computed using the BerkeleyGW code.
\cite{BGW.Deslippe,PRB.Rohlfing,PRB.Hybertsen}
For the unit cell MoS$_2$ calculation, we use a k-point sampling of
24$\times$24$\times$1 and 8,400 valence and conduction states.
We find the quasiparticle band gap of monolayer
MoS$_2$ to be 2.7 eV, which is in
good agreement with previous calculations and experimental measurements.
\cite{PRL.Qiu,PRL.Erratum.Qiu,NL.Krane,JPCC_Thyg}
For the super cell calculations, we use a k-point sampling of 2$\times$2$\times$1 and
19,000 valence and conduction states. 
We find that these parameters are
sufficient to converge the gap at the $K$ point in the Brillouin zone to within 0.2 eV.
%We generate the 19,000 wavefunctions in the following manner. 
The first 9,000 lowest energy wavefunctions are treated using a
plane-wave energy cut off of 150 Ry, and the rest 10,000 wavefunctions 
are treated using a smaller cut off of 50 Ry. We find that this 
method produces accurate results at significantly reduced
computation time. 
The states with low energy, below the Fermi level, are
spatially localised. 
These gradually become more plane-wave like at higher energies. 
The higher energy wavefunctions can thus
be described with a fewer number of plane-waves. 
We have tested this method using the unit cell MoS$_2$. 
We find that the GW band gap is 
unchanged if we use the first 60 wavefunctions at 150 Ry cut off and 
the rest at 50 Ry cut off.  
It is worthwhile to note that the wavefunctions generated with the 
50 Ry cut off need to be orthonormalized to the 150 Ry cut off wavefunctions.
This is done in the following manner for the defect supercell calculations.
We first
generate 9,000 lowest energy wavefunctions
at an energy cut off of 150 Ry. We then generate 19,000 lowest energy
wavefunctions with an energy cut off of 50 Ry. We compute the overlap of 
these 19,000 with the original 9,000 bands. The bands with a large overlap with the 
original wavefunctions are left out, and the remaining are orthonormalized with the
original. 
\cite{PRL.Samsonidze}
%remove ones among them that have a sufficiently high overlap with the original 9,000 bands.
%We orthonormalize the remaining bands with the original 9,000 to
%generate a total of 19,000 wavefunctions. \cite{PRL.Samsonidze}
We use PRIMME \cite{PRIMME1,PRIMME2} to generate the 9,000 bands at 150 Ry cut off and
ScaLAPACK \cite{ScaLAPACK} exact diagonalization routines to generate the 19,000,
50 Ry cut off wavefunctions. 
Further, the static remainder technique \cite{PRB.Deslippe} is used 
to accelerate convergence 
of the calculation with the number of empty states.
A dielectric cut off  of 35 Ry is used.
The Coulomb interaction along the
out-of-plane direction is truncated for the computation of
dielectric matrix and self energy. \cite{PRB.Ismail}
The dielectric function is extended to finite frequencies using
the Hybertsen-Louie generalized plasmon pole (GPP) model. \cite{PRB.Hybertsen}

The substrates included in our calculations are BN, SiO$_2$, graphene,
bilayer graphene (BLG), trilayer graphene (TLG) and graphite.
The wavefunction cut-off used is 70 Ry for BN and SiO$_2$; and 60 Ry
for graphene, BLG, TLG and graphite calculations. K-point sampling used
for graphene, BLG and TLG is 21$\times$21$\times$1. K-point sampling used
for graphite is 21$\times$21$\times$10. 
K-point sampling used for BN and SiO$_2$ is 15$\times$15$\times$1 and
14$\times$14$\times$14, respectively. For the 2D substrates, the cell dimension
in the out-of-plane direction is chosen to match that of MoS$_2$.
We use the semiempirical Grimme \cite{JCC.Grimme} scheme to
account for the van der Waals interactions
between the layers in heterostructures constructed at 
the DFT level to obtain the interlayer spacings.
We perform calculations on the unit cells of
substrates to obtain their irreducible polarizabilities. 
The k-point sampling in the
polarizability calculations is the same as those for the DFT
calculations. The dielectric cut off used for graphene, BLG, TLG and
graphite is 10 Ry. For BN and SiO$_2$, the dielectric cut off used is
12 Ry and 10 Ry, respectively. The number of unoccupied states for graphene,
BLG, TLG and graphite is 250. For BN and SiO$_2$, the number of states is
600 and 300, respectively. For metallic substrates, the polarizability
at the q-point close to the $\Gamma$ point is computed with a finer
k-point sampling of 80$\times$80$\times$1.

\begin{figure*}
  \centering
  \includegraphics[scale=0.12]{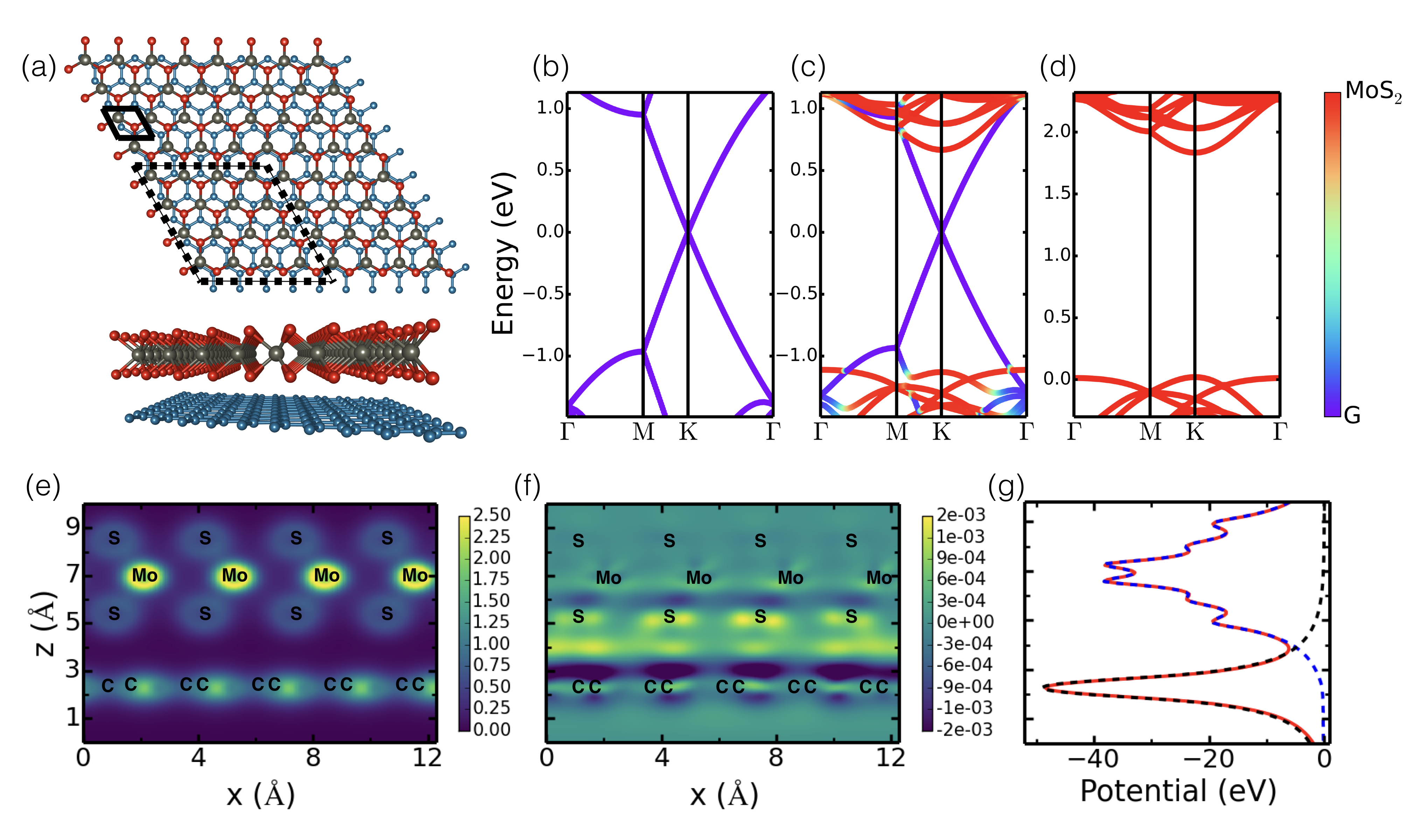}
  \caption
  {\label{fig1}
  (a) Single layer MoS$_2$ on a graphene substrate, top-view and
      side-view. The black solid line
      marks the unit cell of MoS$_2$. The dotted line marks the lattice-matched
      super cell used to perform the DFT calculations.
  (b), (c) and (d)  DFT band structures of free-standing 5$\times$5 super cell of graphene,
      lattice matched graphene-MoS$_2$
      heterostructure and free-standing 4$\times$4 super cell of MoS$_2$, respectively.
      The colors indicate
      the projected weights of the heterostructure wavefunctions onto the free-standing
      layers.
  (e) The charge density of the MoS$_2$-graphene heterostructure,
      in e/$\textrm{\AA}^3$, averaged along one of the in-plane lattice 
      directions.
  (f) The charge density difference, between the MoS$_2$-graphene heterostructure 
      and the corresponding free-standing layers.
  (g) The potential of the MoS$_2$-graphene heterostructure (red line), 
      the free-standing graphene
      and the free-standing MoS$_2$ layer (blue line).
  }
\end{figure*}

\begin{figure}
  \centering
  \includegraphics[scale=0.14]{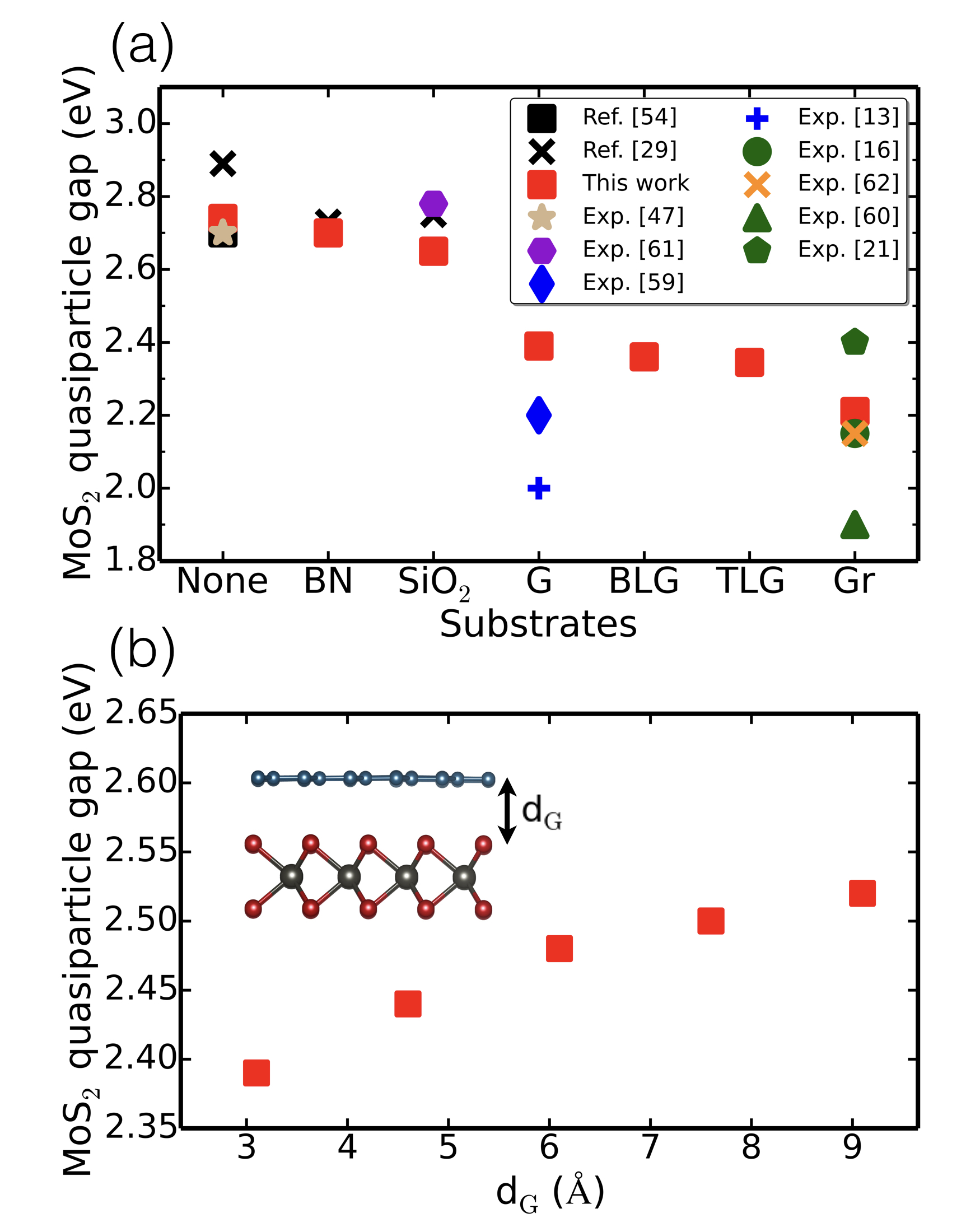}
  \caption
  {\label{fig2}
  (a) Quasiparticle band gap of MoS$_2$, free-standing, and in the presence of monolayer BN,
      bulk SiO$_2$, graphene (G), bilayer graphene (BLG),
      trilayer graphene (TLG) and graphite substrates. Experimental measurements of the
      quasiparticle gap in these systems are also shown in the plot.
  (b) Quasiparticle band gap of MoS$_2$ in the presence of graphene as a function of increasing
      inter-layer spacing, $\mathrm{d}_\mathrm{G}$.
  }
\end{figure}

\section{Effect of substrate screening}

We study the interaction between MoS$_2$ and a substrate at the DFT level by constructing
commensurate super cells that accommodate the two materials with a strain of less than 2\%.
We use a 4$\times$4 super cell of MoS$_2$ and a 5$\times$5 super cell
of graphene.
Fig. \ref{fig1} (a) shows MoS$_2$ on a graphene substrate.
A similar geometry is used for the case of MoS$_2$ on BN since the lattice parameter 
of BN is close to that of graphene. 
The BN or graphene layers are strained to attain a commensurate super cell. 
The relaxed interlayer spacing for the MoS$_2$-graphene heterostructure is 3.1 \AA.
Fig. \ref{fig1} (b) and Fig. \ref{fig1} (d) show the DFT
band structure of the 5$\times$5 super cell of graphene and the 4$\times$4
super cell of MoS$_2$, respectively. Fig. \ref{fig1} (c) shows the band structure of the
MoS$_2$-graphene heterostructure.
The DFT wavefunction of the heterostructure, for a given band and k-point,
has been projected onto the wavefunctions of free-standing graphene and free-standing MoS$_2$.
The projected weights are then portrayed using a color map. Note that the energy of the VBM
has been set to zero in these plots.
It can be seen that interlayer coupling or hybridization
is absent at the VBM and CBM of MoS$_2$ in the heterostructure.
At the DFT level, the band gap of MoS$_2$ in the presence of graphene
is unchanged. This is different from bilayer MoS$_2$ where
the overlap of wavefunctions of similar energies  leads to strong
hybridization and a transition of the gap from direct to indirect \cite{PRB_Naik}.
Slight hybridization is, however, seen far from the Fermi level, leading
to the creation of small gaps in graphene of about 70 meV. These minigaps
have been recently observed in MoS$_2$-graphene heterostructures using
angle-resolved photoemission spectroscopy (ARPES). \cite{NL_Coy,NL_Pier}
Fig. \ref{fig1} (e) plots the charge density of the MoS$_2$-graphene 
heterostructure, $\rho^{\mathrm{MG}}(\mathbf{r})$, averaged along one of 
the in-plane lattice vectors. 
Fig. \ref{fig1} (f) plots the charge density difference, 
$\rho^{\mathrm{MG}}(\mathbf{r}) - \rho^\mathrm{M}(\mathbf{r}) 
- \rho^\mathrm{G}(\mathbf{r}) $, in the same manner.
In the heterostructure, the electronic charge density within each layer 
is slightly rearranged, but there is 
no possibility of charge transfer from one layer to the other due to the 
sizeable energy difference between the graphene Fermi level and the MoS$_2$ CBM.
Our Bader charge analysis further supports the absence of charge transfer. 
%The directionality 
%of the rearrangement of charges is explained by the potential gradient induced 
%in one layer due to the other. 
The directionality
of the rearrangement of charges, leading to the formation of out-of-plane dipole moments,
is explained by the non-uniform
potential gradient induced in one layer due to the other (Fig. \ref{fig1} (g)).
%Fig. \ref{fig1} (g) shows the DFT potential of 
%the individual layers and the heterostructure. 
At the equilibrium spacing, the 
potential from one layer is finite and decreasing in the vicinity of the 
other layer. This gradient acts like an effective electric field for the other 
layer leading to the rearrangement of electrons.

Performing GW calculations on the various super cell geometries 
is computationally demanding.
We instead perform separate unit cell calculations on MoS$_2$ and the substrates.
To take into account the effect of a substrate on MoS$_2$ we 
map the in-plane $\vec{q} + \vec{G}$ vectors of the MoS$_2$ 
irreducible polarizability, $\chi_{\vec{q}}^{\vec{G}\vec{G}'}$, 
to $\vec{q} + \vec{G}$ vectors of the substrate irreducible 
polarizability. The substrate polarizability element corresponding 
to the mapped $\vec{q} + \vec{G}$ vector is then added to the 
polarizability element of MoS$_2$. 
\cite{NMat.Ugeda,NL_Qiu,NL.Aaron}
Using this method,
the band gap reduction is slightly overestimated
for bulk substrates.
Fig. \ref{fig2} (a) shows the quasiparticle band gap of MoS$_2$ in the
free-standing case and in the presence of BN, SiO$_2$, graphene (G), bilayer graphene (BLG),
trilayer graphene (TLG) and graphite (Gr) substrates.
Also marked in the figure are the experimentally measured quasiparticle band
gaps of MoS$_2$ on these substrates.
A significant renormalization to the band gap of MoS$_2$ is captured at the GW level, 
while the gap remains unchanged at the DFT level. 
This is due to the inclusion
of image charge effects at the GW level. 
The more metallic nature of the substrate, larger the 
band gap renormalization. 
A similar trend is observed for molecules 
on a metal substrate, where DFT fails to predict any renormalization to 
the molecular levels. While GW effectively captures 
the non-local screening due to image charge effects and shows a renormalization, in 
agreement with experimental findings. \cite{PRL.Rohlfing,PRL.Neaton, PRL.Thyg}
The renormalization of the MoS$_2$ quasiparticle band gap in the
presence of BN and SiO$_2$ is 40 meV and 90 meV respectively.
In the presence of graphene,
BLG, TLG and graphite; the renormalization is 350 meV, 380 meV, 400 meV and
530 meV respectively. Our result for the renormalization in the presence of graphene
is in good agreement with a recent GW calculation on the explicit MoS$_2$-graphene
heterostructure. \cite{JPCC_Thyg}
The value of the quasiparticle
band gap measured experimentally is in excellent agreement for the free-standing case
and in the presence of SiO$_2$ substrate (Fig. \ref{fig2} (a)). 
The experimental quasiparticle
band gap of MoS$_2$ measured in the presence of graphene and graphite substrate,
on the other hand, is varied and falls in
the range of 1.9 to 2.4 eV (Fig. \ref{fig2} (a)).
In Fig. \ref{fig2} (a), all the experimental values reported are
measured using scanning tunneling spectroscopy except the one 
on SiO$_2$ substrate, which uses photoluminescence excitation 
spectroscopy.
We additionally study the effect of MoS$_2$-graphene interlayer spacing
on the quasiparticle band gap of MoS$_2$. We find that 
the gap is sensitive to the spacing and
can be tuned from 2.4 eV to 2.5 eV (Fig. \ref{fig2} (b)).
We estimate the error in our calculation of the quasiparticle 
band gap is 100 meV in the case 
the 2D substrates and 150 meV in the case of bulk substrates.
There exist other factors in the experiment that could 
lead to further renormalization of the 
band gap in MoS$_2$. These include the effect of 
carrier-induced plasmons in the system, which
have recently been shown to close the gap by upto 150 meV. \cite{PRL_Liang,PRL.Yao}
Additional screening from the metallic tip of the scanning tunneling microscope could
also further renormalize the band gap. \cite{CMS.Georgy}

\begin{figure}
  \centering
  \includegraphics[scale=0.08]{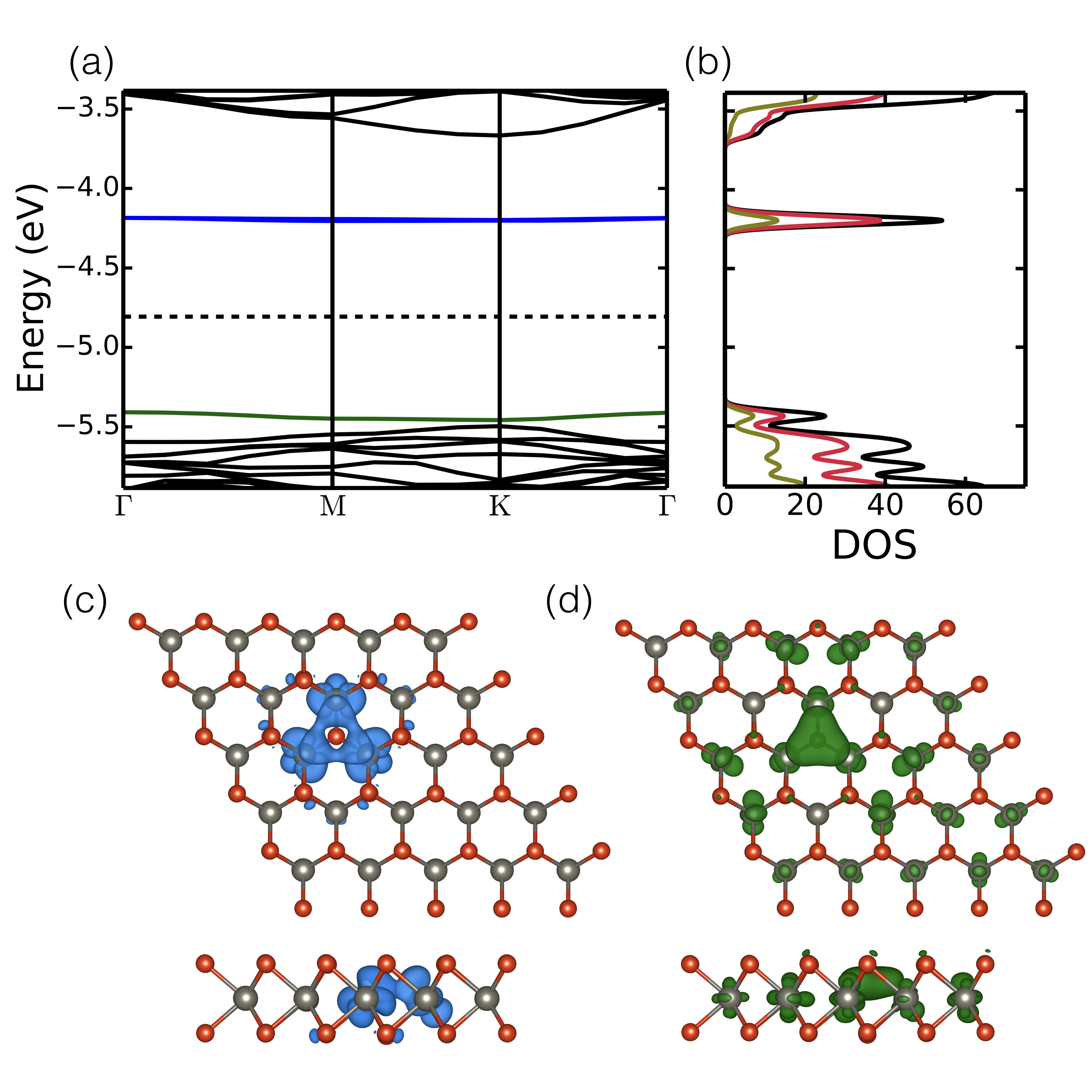}
  \caption
  {\label{fig3}
  (a) DFT computed band structure of a sulfur vacancy defect in a 5$\times$5 super cell of MoS$_2$. Three
      defect states are induced in the gap. The filled defect level 
      is indicated in green, the unfilled
      levels are doubly degenerate and indicated in blue. 
      The black dashed line marks the Fermi level.
  (b) Partial density of states (DOS) of the 5$\times$5 super cell of MoS$_2$ with a 
      sulfur vacancy. The red line shows the Mo-$d$ contribution and the green 
      line shows the S-$p$ contribution to the total density of states (black).
  (c) and (d) Isosurface of the defect levels induced in 
      the gap of MoS$_2$. The wavefunction plotted in
      blue is the corresponding unfilled defect level in 
       the band structure and the one plotted
      in green is the filled defect level. The top view as well as the side view are shown.  
}
\end{figure}

\section{Sulfur vacancy defect}
\begin{figure}
  \centering
  \includegraphics[scale=0.08]{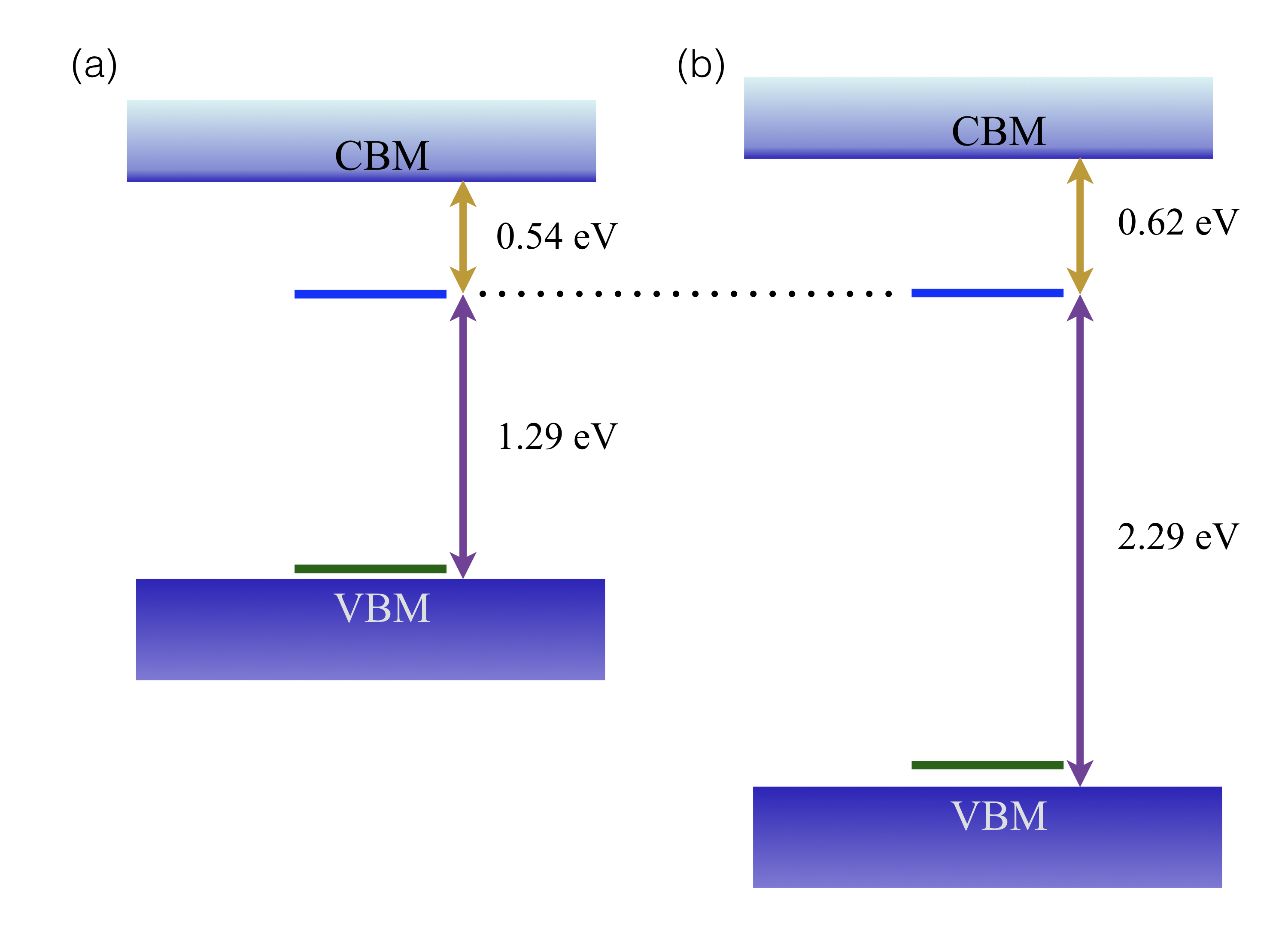}
  \caption
  {\label{fig4}
  (a) Schematic of the DFT prisitine VBM, pristine CBM and the defect levels, plotted with respect to the vacuum level.
  (b) Schematic of the GW pristine VBM, pristine CBM and the defect levels, plotted with respect to the vacuum level.
      The dotted line is a guide to the eye, showing that the unfilled defect levels line up between DFT and
      GW levels of theory.
  }
\end{figure}

Fig. \ref{fig3} (a) shows the DFT band structure of 5$\times$5 super cell of MoS$_2$ with a sulfur vacancy defect.
Three defect states are induced in the gap on introducing the vacancy: one filled (indicated by green)
bonding state and two degenerate unfilled (indicated by blue) anti-bonding states. 
The charge density associated with these
defect states is shown in Fig. \ref{fig3} (c) and (d).
The empty states are localised over a smaller region in the
material as compared to the filled state.
These defect states are dominantly of the Mo-$d$ character
(Fig. \ref{fig3} (b)).
We compare the VBM and CBM of the pristine MoS$_2$ system, and the defect levels
with respect to the vacuum level as computed within DFT and GW. Fig. \ref{fig4} (a) and (b) 
shows a schematic of this
comparison. We find that the DFT calculated CBM and the GW calculated 
CBM differ by about 0.1 eV, while the respective
VBMs are different by 1 eV. 
Interestingly, the empty defect levels are found to line up. 
The filled defect level on the
other hand remains shallow and close to the VBM.
It has been shown that the CTLs of bulk systems line up between 
DFT and GW with 
respect to the average electrostatic potential in the system. 
\cite{JPCM_Chen}
Here, we find that the defect levels line up with respect to the 
vacuum level, while the CTLs do not
(Fig. \ref{fig4}).

\begin{table}
\centering
\begin{tabular}{c@{\hskip 0.07in}r@{\hskip 0.07in}r@{\hskip 0.07in}r@{\hskip 0.07in}r}
\hline
\hline
BZ sampling& \multicolumn{1}{c@{\hskip 0.07in}}{24$\times$24$\times$1} & \multicolumn{1}{c@{\hskip 0.07in}}{12$\times$12$\times$1} & \multicolumn{1}{c@{\hskip 0.07in}}{12$\times$12$\times$1} & \multicolumn{1}{c@{\hskip 0.07in}}{10$\times$10$\times$1}
\\
%BZ sampling& \multicolumn{1}{c@{\hskip 0.07in}}{24$\times$24$\times$1} & \multicolumn{1}{c@{\hskip 0.07in}}{12$\times$12$\times$1} & \multicolumn{1}{c@{\hskip 0.07in}}{12$\times$12$\times$1}

N$_b$ & \multicolumn{1}{r@{\hskip 0.07in}}{6000} & \multicolumn{1}{r@{\hskip 0.07in}}{6000} & \multicolumn{1}{r@{\hskip 0.07in}}{750} & \multicolumn{1}{r@{\hskip 0.07in}}{750}
         \\
\hline
Gap at K (eV) & 2.74 & 2.79 & 2.78 & 2.85 \\
Gap at $\Gamma$ (eV) & 4.02 & 4.06 & 4.08 & 4.13  \\
\hline
%\multicolumn{1}{c}{$\Delta\Sigma$} & & & & & & \\
%(24$\times$24$\times$1)  & & & & & &\\
%&8.3

\hline
\hline
\end{tabular}
\caption{ Convergence of the band gap as a function of Brillouin zone sampling and
number of bands.
}
\end{table}

\begin{figure}
  \centering
  \includegraphics[scale=0.3]{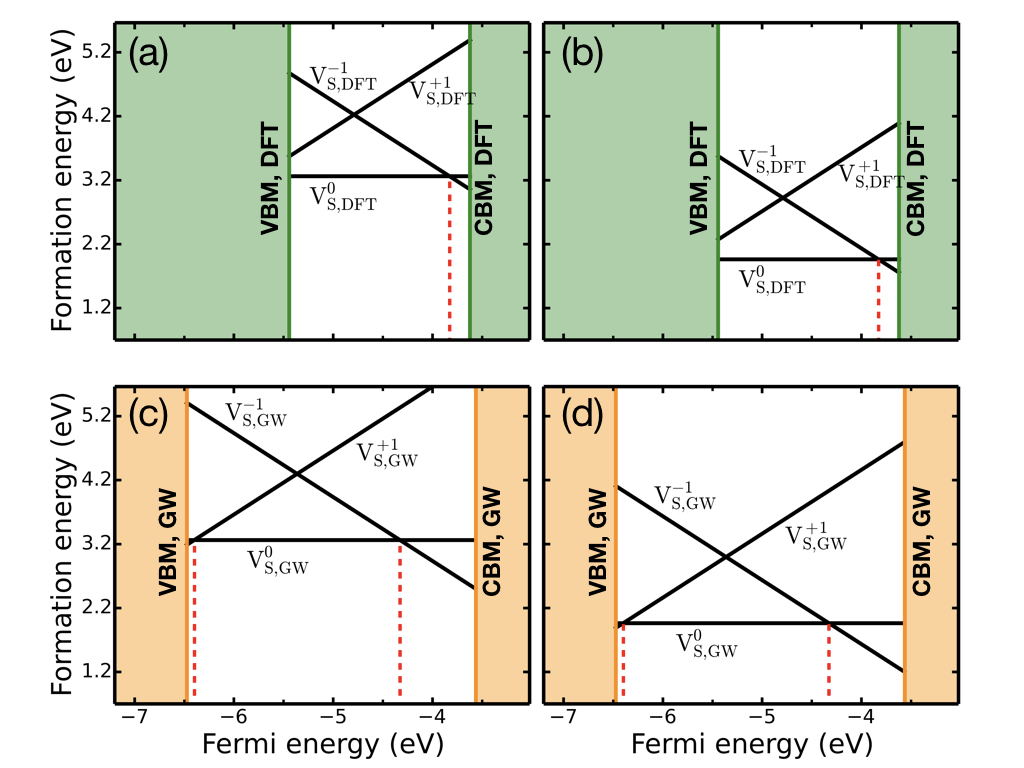}
  \caption
  {\label{fig5}
  Formation energy of the sulfur vacancy in different charge states as a function of the Fermi level.
  The Fermi level is taken to scan the energy range between the pristine VBM and CBM.
  (a) and (b) Computed at the DFT level, for sulphur rich and sulphur poor conditions respectively.
  (c) and (d) Computed using the DFT+GW formalism, for sulphur rich and sulphur poor conditions respectively.
  The charge transition levels that appear in the gap are marked with red dashed lines.
  }
\end{figure}

The formation energy of a sulfur vacancy in charge state $q$ is given by,
\begin{multline}
 \label{eqn1}
  \mathrm{E}_{q}^{f}[\vec{\mathrm{R}}_{q}](E_{F}) = 
  \{\mathrm{E}^{\mathrm{tot}}_{q}[\vec{\mathrm{R}}_{q}] + \mathrm{E}^{\mathrm{corr}}_q \} 
  - \mathrm{E}_{\mathrm{pristine}}   \\
  + q \{ \epsilon_{\mathrm{vbm}}^{\mathrm{pristine}}+
  E_{F} - \Delta V_{0/p} \} - \mu_S
\end{multline}
where $\mathrm{E}^{\mathrm{tot}}_{q}[\vec{\mathrm{R}}_{q}]$ refers to the total energy of
the 5$\times$5 super cell of MoS$_2$ containing the defect in charge state $q$. $\vec{\mathrm{R}}_{q}$
refers to the relaxed atom positions in the super cell of the defect system in charge state q.
$\mathrm{E}^{\mathrm{corr}}_q$
is the electrostatic correction term to account for the spurious interaction of the charged defect with
its periodic images. This term is zero for the case of the neutral defect. $\mathrm{E}_{\mathrm{pristine}}$
is the total energy of a pristine super cell of MoS$_2$ of the same size. The formation energy is a function
of the Fermi level with respect to the VBM of the pristine system, $\epsilon_{\mathrm{vbm}}^{\mathrm{pristine}}+
E_{F}$. $\Delta V_{0/p}$ is the potential alignment term found by comparing the elecrostatic potential
of the neutral defect cell and pristine cell, far from the defect. $\mu_S$ is the chemical potential of the
sulfur atom removed from the pristine system to form the vacancy defect. This reference can be chosen to simulate
sulfur-rich or sulfur-poor ambient conditions. For the sulfur-rich conditions, the chemical potential is chosen
from the cyclo-S$_8$ allotrope of sulfur. For sulfur-poor conditions, the chemical potential is 
chosen 1.3 eV below, the potential at which MoS$_2$ is reduced to body-centered cubic (BCC) Mo metal 
\cite{JC.Schweiger,PRB.Norskov}. 
Fig. \ref{fig5} (a) and (b) plot the DFT computed formation energy
of the sulfur vacancy in $0$, $-1$ and $+1$ charge states as a function of the Fermi level. The Fermi level scans
the pristine MoS$_2$ gap. The formation energy here is determined following Eqn. \ref{eqn1} using
the DFT computed total energy differences. The electrostatic correction term is determined to be $0.1q^2$ eV,
where $q$ is the charge state of the vacancy.
Charge transition level, the Fermi level at which the formation energy of
one charge state of the defect is equal to that of another is given by:

\begin{equation}
  \label{eqn2}
  \varepsilon^{\textrm{q}/\textrm{q-1}} = \textrm{E}_{\textrm{q-1}}^f[\vec{\textrm{R}}_{\textrm{q-1}}] 
  (E_{\textrm{F}}=0)  - \textrm{E}_{\textrm{q}}^f[\vec{\textrm{R}}_{\textrm{q}}] (E_{\textrm{F}}=0)
\end{equation}
The only charge transition level stable in the gap at the DFT level
is $\varepsilon^{0/-1}$ = 1.6 eV from the VBM. This is in good agreement with previous
calculations \cite{PRB_Noh,PRB_Komsa2}.

Within the DFT+GW formalism, the expression for the charge transition level 
can be rewritten into two parts.
One that involves adding an electron to the system, and the other that takes into account
the lattice relaxation effects due to to the added electron \cite{PRL_Jain}. 
The former is evaluated as
a quasiparticle excitation at the GW level and the latter is evaluated at the DFT level.
\begin{equation}
  \label{eqn3}
  \begin{split} 
  \varepsilon^{\textrm{q}/\textrm{q-1}} &= (\textrm{E}_{\textrm{q-1}}^f[\vec{\textrm{R}}_{\textrm{q-1}}] 
  - \textrm{E}_{\textrm{q-1}}^f[\vec{\textrm{R}}_{\textrm{q}}] )
  + ( \textrm{E}_{\textrm{q-1}}^f[\vec{\textrm{R}}_{\textrm{q}}]
     - \textrm{E}_{\textrm{q}}^f[\vec{\textrm{R}}_{\textrm{q}}] ) \\
  &= \textrm{E}_{\textrm{relax}} + \textrm{E}_{\textrm{QP}}
  \end{split} 
\end{equation}
For the $\varepsilon^{\textrm{0}/\textrm{-1}}$ evaluated using the DFT+GW formalism,
we find $\textrm{E}_{\textrm{QP}}$ = 2.3 eV and $\textrm{E}_{\textrm{relax}}$ = -0.1 eV. The charge transition level
is hence 2.2 eV above the pristine VBM. For the $\varepsilon^{\textrm{+1}/\textrm{0}}$, we find
$\textrm{E}_{\textrm{QP}}$ = 0.1 eV and $\textrm{E}_{\textrm{relax}}$ = -0.01 eV, giving the charge transition level
$\sim$ 0.1 eV above the VBM. Fig. \ref{fig5} (c) and (d) show the plot of formation energy with respect to the Fermi
level computed using the DFT+GW formalism. Note that we do not add any electrostatic
correction terms here since the quasiparticle excitation energies are taken from the neutral system.
The $\varepsilon^{\textrm{0}/\textrm{-1}}$ computed using hybrid functionals in the literature
are 1.9 eV \cite{PRB_Noh} and 1.6 eV \cite{PRB_Komsa2} above the VBM. The $\varepsilon^{\textrm{+1}/\textrm{0}}$
computed using
hybrid functionals is found to be below the VBM.

\section{Substrate screening effects on the CTLs}
\begin{figure}
  \centering
  \includegraphics[scale=0.12]{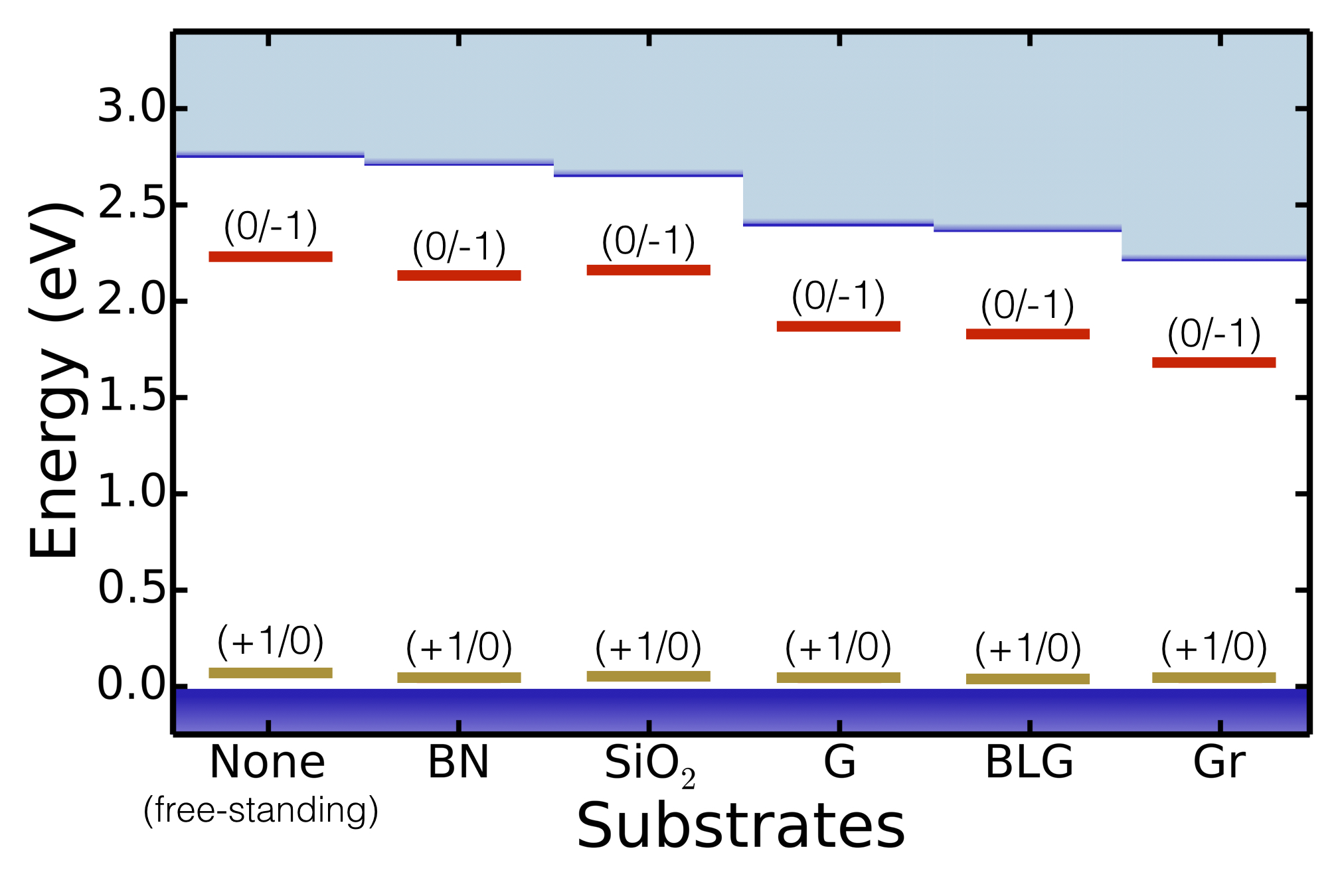}
  \caption
  {\label{fig6}
   The $\varepsilon^{\textrm{+1}/\textrm{0}}$ and $\varepsilon^{\textrm{0}/\textrm{-1}}$
   charge transition levels of the sulfur vacancy defect computed
   using the DFT+GW formalism. The levels are shown with respect to the valence band maximum
   of pristine MoS$_2$ in the presence of BN, silica, graphene (G), bilayer graphene (BLG) and
   graphite (Gr) substrates.
  }
\end{figure}

The presence of substrates leads to a renormalization of the pristine quasiparticle band gap in
MoS$_2$ (Fig. \ref{fig2} (a)), as well as the term 
$\textrm{E}_{\textrm{QP}}$ in Eqn \ref{eqn3}
for the CTL. We compute the renormalization to $\textrm{E}_{\textrm{QP}}$
from the super cell calculation. The renormalization to the pristine band gap,
on the other hand, is taken from the unit cell calculations (Fig. \ref{fig2} (a)).
We also assume that the $\textrm{E}_{\textrm{relax}}$ term is the same in the presence
and absence of substrates. Fig. \ref{fig6} shows the CTLs in the quasiparticle band
gap of pristine MoS$_2$ for the various substrates. 
The $\varepsilon^{\textrm{+1}/\textrm{0}}$
is pinned close to the VBM, within 100 meV. 
The defect level involved in this transition is a relatively shallow level with a 
larger band width (Fig. \ref{fig3} (a)). The larger band width indicates slight hybridization with 
the valence band edge. Hence the effect of substrate screening on this level is similar to that 
of the VBM. This leads to the pinning of $\varepsilon^{\textrm{+1}/\textrm{0}}$.
The
$\varepsilon^{\textrm{0}/\textrm{-1}}$, with
respect to the VBM, is renormalized by about the same amount as the band gap.
The anti-bonding character of the empty defect states is similar to that of the
CBM of the pristine MoS$_2$ system. \cite{PRB.Sensoy} 
The $\varepsilon^{\textrm{0}/\textrm{-1}}$ is thus pinned about 500 meV below the CBM in the
presence of substrates as well as in the free-standing configuration
(Fig. \ref{fig6}).

\section{Conclusion}
We have studied the effect of substrate screening on the electronic structure of
monolayer MoS$_2$. The substrates included in our calculations are: BN, SiO$_2$, graphene,
bilayer graphene and graphite. These substrates lead to a significant
renormalization of the quasiparticle
band gap of MoS$_2$. In the presence of graphene and graphite substrates, in particular,
we find a large reduction of 350 meV and 530 meV, respectively.
These results are in good agreement with recent experimental measurements
on these systems. \cite{NC.Song}
We have also studied the charge transition levels of
sulfur vacancy defects in MoS$_2$ using the DFT+GW formalism. We find two
CTLs lying in the pristine quasiparticle band gap of MoS$_2$,
the (+1/0) and the (0/-1) level. The (+1/0) level and (0/-1) level are found 0.07 eV and 2.14 eV above
the pristine VBM, respectively. 
The stability of the -1 charge state is in good agreement with recent experimental 
findings. 
We also compute the CTLs in the presence of
substrates. 
The CTLs show a renormalization similar to that of pristine MoS$_2$ and
\emph{remain} in the pristine band gap of MoS$_2$.
With respect to the VBM, the (0/-1) level is renormalized
by the same amount as the gap. The (0/-1) level is thus pinned about 500 meV below the CBM
for free-standing MoS$_2$ case as well as in the presence of substrates.
The (+1/0) level, on the other hand, lies less than 100 meV above the VBM in all the cases.
The tuning of the defect levels with choice of substrate would aid in 
tuning the binding of hydrogen at sulfur vacancy sites, which is important 
to optimize the hydrogen evolution reaction. Charged-defect scattering from 
the -1 charged sulfur vacancy can be avoided if the Fermi level of the system is below
the computed CTL. This could improve the mobility of carriers in MoS$_2$. 
The possibility of tuning the CTLs with choice of substrate need not be 
restricted to MoS$_2$.  Other transition metal dichalcogenides and two-dimensional 
materials could also be expected to show a 
similar tuning of the defect CTLs. 

% If you have acknowledgments, this puts in the proper section head.
\begin{acknowledgments}
We thank the Supercomputer Education and Research Centre (SERC) at IISc
for providing the computational facilities. Computational facilities from
C-DAC's PARAM Yuva II were also used in this work.
We thank Felipe H. da Jornada and Diana Y. Qiu for discussions.
\end{acknowledgments}

% Create the reference section using BibTeX:
%\bibliography{substratescreening.bib}
%merlin.mbs apsrev4-1.bst 2010-07-25 4.21a (PWD, AO, DPC) hacked
%Control: key (0)
%Control: author (8) initials jnrlst
%Control: editor formatted (1) identically to author
%Control: production of article title (-1) disabled
%Control: page (0) single
%Control: year (1) truncated
%Control: production of eprint (0) enabled
%
%merlin.mbs apsrev4-1.bst 2010-07-25 4.21a (PWD, AO, DPC) hacked
%Control: key (0)
%Control: author (8) initials jnrlst
%Control: editor formatted (1) identically to author
%Control: production of article title (-1) disabled
%Control: page (0) single
%Control: year (1) truncated
%Control: production of eprint (0) enabled

\end{document}